
\input harvmac

\Title{hep-th*** RU-95-88}
{\vbox{\centerline{The Number of
States of}
\centerline{ Two Dimensional Critical String Theory}}}
\bigskip
\centerline{Tom Banks  }
\smallskip
\centerline{\it Department of Physics and Astronomy}
\centerline{\it Rutgers University, Piscataway, NJ 08855-0849}
\smallskip
\centerline{Leonard Susskind}
\smallskip
\centerline{\it Physics Department}
\centerline{\it Stanford University, Stanford, CA 94305}
\bigskip
\baselineskip 18pt
\noindent

We discuss string theory vacua which have the wrong number of spacetime
dimensions, and give a crude argument that vacua with more than four
large dimensions are improbable.  We then turn to two dimensional vacua,
which naively appear to violate Bekenstein's entropy principle.  A
classical analysis shows that the naive perturbative counting of states
is unjustified.  All excited states of the system have strong coupling
singularities which prevent us from concluding that they really exist.
A speculative interpretation of the classical solutions suggests only a
finite number of states will be found in regions bounded by a finite
area.  We also argue that the vacuum degeneracy of two dimensional
classical string theory is removed in quantum mechanics.  The system
appears to be in a Kosterlitz-Thouless phase.  This leads to the
conclusion that it is also improbable to have only two large spacetime
dimensions in string theory.  However, we note that, unlike our argument
for high dimensions, our conclusions
about the ground state have neglected two dimensional quantum
gravitational effects, and are at best incomplete.

\Date{11/95}

\newsec{Introduction}

String theory appears to have many classical ground states which bear no
resemblance whatsoever to the real world.  In particular, many of them
have the wrong number of large spacetime dimensions.  What is perhaps
more disturbing, is that many of these wrong dimensional vacua have a high
degree of supersymmetry.  Very strong symmetry arguments then suggest
that they are exact, stable quantum ground states of the system.
Why then are they not realized in the world of phenomena?

For ground states of high spacetime dimension, the beginnings of an
argument can be constructed.  Imagine, following Einstein, that there is
a general principle that only compact spatial dimensions need to be
considered.  Then the idea of decoherence of states with different
expectation values of homogeneous scalar fields is only an approximate
one.  Instead of a moduli space of vacua we would have a ground state
wave function which is a function on moduli space.  In a low energy,
``minisuperspace'' approximation (and neglecting for the moment the
coupling to gravity) it would be the ground state
of the nonlinear quantum mechanics
model on moduli space defined by the tree level string
effective action.  Apart from gravitational effects,
corrections to this approximation can be controlled by
constructing an effective action for the homogeneous modes by
integrating
out string modes with finite mass and momentum. The corrections
are negligible near the
boundaries of moduli space where more than four dimensions are large.
Indeed, in high dimension this is true even for gravitational effects.
In particular, infrared instabilities which could lead to a potential on
moduli space, and perhaps inflation, vanish in the limit in which the
system has more than four noncompact dimensions.
When the theory has a sufficiently large supersymmetry, they are
completely absent.   We conclude that the dynamics of the homogeneous
modes is well approximated in this limit by the quantum
mechanics of free motion on a noncompact space of finite
volume\ref\moorehorne{J.Horne, G.Moore, {\it Nucl. Phys.}{\bf
B342},(1994),109, hep-th/9403058}.
The ground state wave function is a constant
and the probability of being at any particular point of moduli space
is determined by the Zamolodchikov volume element.  {\it The probability
of radial moduli being much larger than the string scale, vanishes like
a power of the radius.}

When we consider the coupling of the system to gravity, this reasoning
can be falsified by the phenomenon of inflation.  We will not go into a
detailed discussion of how quantum string dynamics in four dimensions might
lead to inflation, but one thing is certain.  In order to have inflation
we must have a potential on moduli space. Furthermore, the natural scale
of the potential must be much smaller than the Planck mass (this is the
reason why fields other than moduli are not good candidates for the
inflaton).  Planck scale inflation leads to density fluctuations of
order one when it ends.  A candidate inflationary universe in such a
situation would become highly inhomogeneous and collapse into a
collection of black holes on microscopic time scales.  A successful slow roll
inflationary scenario requires the potential to remain much smaller than the
Planck scale while the fields vary by amounts of order the Planck scale.

 In dimension higher than $4$
there does not appear to be any infrared dynamics which could generate
a potential with natural height smaller than the Planck scale.  For
vacua with a high degree of SUSY, the same arguments which prove their
stability show that {\it no} potential on moduli space is possible.
Thus the probability of finding oneself in such a state would be
determined by the measure on moduli space, and string theory could be
said to predict that it is highly improbable to
find the world in a state where more than four dimensions are
large.

Note that the key point distinguishing four dimensions
here is the ability of marginally
relevant four dimensional gauge couplings to produce a large hierarchy
of scales.  The small potential on moduli space is determined by
nonperturbative gauge dynamics.
It has often been speculated that this property of four
dimensional gauge theories would be an important part of the answer to
why we live in four dimensions.  Here we have used four dimensional
gauge dynamics to construct an explicit
cosmological mechanism for explaining this fact.
\foot{It is an interesting open question whether this reasoning
also rules out vacua with four large dimensions and extended SUSY.  The
answer to this has to do with whether such vacua are continuously
connected (in configuration space, not along manifolds of classical
solutions\ref\dixon{T.Banks, L.Dixon, {\it Nucl. Phys.}{\it
B307},(1988),93.}) to vacua where potentials can be generated.
Inflation could then occur in regions where extended SUSY is broken, but
the system could settle down into a vacuum with extended SUSY.}

Vacua with fewer than four dimensions are another question entirely.
The above argument does not immediately apply to them.
Of particular concern, are two dimensional vacuum states, because they appear
to violate the Bekenstein bound on the entropy of a system of given
energy.
In $D$ spacetime dimensions, the bound says that the region enclosed
inside a $D-2$ surface can have a number of degrees of freedom
 no larger than the
$D-2$ volume of the surface in Planck units.  A two dimensional
compactification of string theory should be viewed as a ten
dimensional spacetime with all but one of the spatial directions
compactified at the string scale.  If we consider the region to the left
of
some point in the large direction, it is bounded by a surface of volume
the string scale.  For any fixed value of the dilaton, Bekenstein's
principle would predict that the region had a finite number of states.
But string theory appears to predict that the low energy dynamics in
this vacuum state is given by a two dimensional field theory, and one
would imagine that this implied that there were an infinite number of
states to the left of a given point.

In this note, we will show that this expectation is incorrect, and that
in fact string theory gives no clear prediction of the number of states.
Usually, one establishes the existence of excitations of a given ground
state by studying small fluctuations around a classical vacuum
configuration, and establishing the spectrum of states in a systematic
perturbation expansion.  In two dimensions, as a consequence of the
infrared divergences of long range fields\foot{We note that the vacua in
question are Lorentz invariant and have no linear dilaton condensate.
Consequently the graviton and dilaton fields are long ranged.},
it is not quite so obvious how to proceed.

We will study the nonlinear classical field equations of the low energy
effective field theory.  We show that unless the string coupling is
exactly zero, arbitrarily small, arbitrarily
smooth waves of incoming massless moduli fields lead to naked
singularities at which the coupling goes to infinity at finite points of
space time.  Indeed, these singularities are a consequence of the
constraint equations, and follow along with the motion of the massless
waves.  There is no regular classical solution describing low amplitude
smooth coherent states of moduli.  We suggest that this is evidence
for the nonexistence of the naive spectrum of excitations of two
dimensional string vacua.  This would remove the apparent contradiction
with Bekenstein's principle, and neatly explain why the stringy world is
not two dimensional\foot{Any Lorentz invariant vacuum state is only an
idealization of the situation in the real world.  The universe has
undergone a complicated evolution, and only approximately resembles
empty Minkowski space, at late times.
A vacuum state with no nearby excitations can
never be the endpoint of cosmological evolution.}.

One possible interpretation of the singularities which we have discovered
is the notion of \lq\lq spontaneous contraction'' of two dimensional space
time.  That is, we interpret the singularity of the classical solution
as defining a boundary of space, (making it semi-infinite rather than
infinite).  The proper distance to the boundary is finite, so if this
interpretation is correct, we should expect that there are only a finite
number of states between any finite point and the boundary.  This would
remove the contradiction with Bekenstein's principle.  Of course, since
the region near the boundary is strongly coupled, we cannot give a
convincing argument for this interpretation on the basis of classical
field equations.  Also note that these boundaries are lightlike, which
makes their interpretation somewhat obscure.

In passing, we note also that the {\it vacuum} state of two dimensional string
theory is quite different than that suggested by classical reasoning
.  Standard arguments suggest that two dimensional string theory has a
large moduli space parametrized by a noncompact coset space.  It is well
known however that nonlinear models in two dimensions have
a unique vacuum.  In other words, as in the nonlinear quantum mechanics
we encountered in higher dimensional cosmology, the ground state
wave functional of
the model is spread over the entire moduli space.  We do not know how to
calculate the wave functional, (expecially in string theory where we
must also solve two dimensional gravity) but the classical picture of a moduli
space of vacuum states is likely to be misleading.

\newsec{Classical Gravity in Two Dimensions}

We utilize the notation of Sen's paper\ref\sen{A.Sen, {\it Nucl.
Phys.}{\bf B447},(1995),62, hep.th/9503057.} to write the equations
of two dimensional string theory in conformal gauge as:
\eqn\one{\partial_+ \partial_- (e^{-\Phi}) = 0.}
\eqn\two{\partial_{\pm} (ln \lambda)\partial_{\pm} (e^{-\Phi}) =
\partial^2_{\pm} (e^{-\Phi}) + {1\over 4} e^{-\Phi} Tr (\partial_{\pm} M
L \partial_{\pm} M L).}
\eqn\three{G_{\mu\nu} = \lambda e^{2\Phi}\eta_{\mu\nu}.}
\eqn\four{x^{\pm} = {1\over \sqrt{2}} (x^0 \pm x^1).}

The moduli are incorporated in the matrix $M$, which parametrizes a
noncompact coset space.
We have changed a typographical sign error in Sen's paper.  The kinetic
terms of the dilaton and moduli fields should have opposite signs.
The matrix $L$ is described in \sen .  We will not need its
detailed form.  Indeed, all we really need to use is the fact that the
lagrangian of the moduli is conformally invariant, and that the far
right hand sides of \two are the left moving and right
moving stress tensors of the moduli.  It follows that $ln \lambda$ is the sum
of a function of $x^+$ and a function of $x^-$ and can therefore be
eliminated by further choice of gauge.  The spacetime metric is
therefore flat, and the inverse square of the
string coupling $e^{-\Phi}$ is the sum of a left
moving and right moving function.  Let us assume first that we have only
left moving stress energy, which is a smooth function of $x^+$,
vanishing at infinity.

Before the left moving pulse arrives, the system is in its vacuum state,
and the dilaton is constant.  This boundary condition leads to the
integral equation:
\eqn\inteq{e^{-\Phi} = e^{-\Phi_0} -
\int_{-\infty}^{x^+}\int_{-\infty}^{y} dy dz T_{++} (z).}
But now notice that, if the left moving flux is positive, the integral
grows without bound as $x^+ \rightarrow\infty$.  Therefore, unless the
initial value of $e^{-\Phi}$ is infinite ({\it i.e.} the coupling is
zero), it will become negative at some finite $x^+$.  This is
inconsistent
with the reality of the dilaton.  Before the negative region is reached,
the string coupling goes to infinity.

Thus, in two dimensional string theory, any localized
coherent pulse of positive energy matter, no matter how small
and smooth, carries with it a singular region in which the semiclassical
expansion breaks down.  There is no sensible semiclassical picture of
excited states of a two dimensional string vacuum.  We cannot of course
prove that string quantum mechanics does not solve this problem, and
restore the naive perturbative picture of an infinite number of states
in a region bounded by a finite area, but there is no particular reason
to think that this is so.

One possible interpretation of the solutions which we have found, is
that of regular
spacetimes with a lightlike boundary along the trajectory where the
coupling reaches infinity.  The proper distance to this boundary is
finite, so even in (cutoff) field theory, we expect to be able to associate
only a finite number of quantum states with the region between any
finite point of space and the boundary\foot{Let us note here an
interesting interpretation (due to E.Witten) of the claim that there are
only a finite number of quantum states.  Two dimensional string theory
has an enormous duality group, which acts on its phase space.  Perhaps
the fundamental domain of this group action on phase space has finite
volume. Then the quantum system would have a finite number of states.}.
This interpretation thus
removes the contradiction with Bekenstein's principle.

Alternatively, it might simply be that the vacuum has {\it no} excited
states.  In this view, the classical singularity is a signal of a
sickness in the quantum theory.  We are not sympathetic to this point of
view.  We prefer to believe that string theory is a sensible nonsingular
theory even in two dimensions.  Indeed, as explained in the
introduction, one should probably think of all vacuum states of string
theory (at least with the same number of supersymmetries) as living in a
single connected configuration space.  A disease of the two dimensional
vacuum should be explained either in the innocuous manner of the
previous paragraph or as an instability that drives the system into
regions where the low energy two dimensional lagrangian no longer serves
as a good approximation to string theory.  To examine this question
further we will study the two dimensional vacuum state.

\subsec{The Two Dimensional Vacuum State}

We have argued that there is much less than meets the perturbative eye
in the excitation spectrum of two dimensional string vacua.  We also
believe that the perturbative picture of a moduli space of vacua is
substantially modified in the quantum theory.  Of course, the enormous
duality group of two dimensional string theory already restricts
the naive space of vacua to the fundamental domain of the modular group.
However, the fundamental domain of the group's action on constant fields is
still a multidimensional, noncompact space of finite volume.

In flat spacetime two dimensional field theory, nonlinear models with
noncompact target space and negative curvature, are infrared stable.
They are well defined as quantum field theories only in the presence of a
cutoff.  For finite cutoff, such as that provided by string theory, they
are interacting quantum mechanical theories.  The expansion around
a given point in the moduli space is plagued with infrared divergences.
Infrared stability implies that the long distance physics is
described by a Kosterlitz Thouless phase.  That is, the fluctuations
around a point are given by logarithmic corrections to a Gaussian model.
The long distance observer sees a flat metric on moduli
space, since the system is infrared free.
Thus, one would imagine that fluctuations tend to drive one into
the region of large moduli.

In string theory however these predictions are modified by the effect of
two dimensional gravity and by the fact that the region of large moduli
is really higher dimensional.  We have seen that the gravitational
modifications are dramatic for states of finite energy, but we do not
know how to compute
the gravitational effects on the vacuum.  There is however another
important modification of the picture outlined above.  When the moduli
get large, large numbers of light states descend from the string scale.
They make the theory higher dimensional (at least in the
noncompact parts of moduli space which we understand), and cut off the
two dimensional fluctuations.  Thus the conclusion that one is driven to
large moduli is invalid.

Indeed, it would seem that we have confused the issue by decompactifying
the last spatial dimension.  If it is compact, we go back to the
discussion of the introduction and find that the region of large moduli
is not favored.  Our discussion of the low energy infrared stable
nonlinear model suggests that no potential will be generated in the part
of moduli space in which there is one large spatial dimension.
Note that this did not depend on the large number of supersymmetries in
the problem.  The low energy limit of all two dimensional
compactifications, whether supersymmetric or not, has a moduli space
with negative Ricci curvature.

The dynamics of two dimensional gauge fields, which we have hitherto
neglected, does not seem to change this conclusion.  Gauge couplings are
dimensionful in two spacetime dimensions, and their order of magnitude
is given by the string scale.  The long distance fluctuations of a two
dimensional gauge theory are completely described by a conformally
invariant current algebra of gauge invariant currents.  This does not
generate a potential on moduli space which could alter the conclusion
that large two dimensional volume is improbable.

The most serious loophole in these arguments is our neglect of two
dimensional gravitational quantum fluctuations.  We have seen that
classical gravity dominates the physics of classical states other than
the vacuum and completely violates our flat space intuitions.  It would
be rather remarkable if quantum gravity
did not also modify our picture of the vacuum structure. Unfortunately,
we do not at present know how to estimate quantum gravitational effects.

To summarize, classical singularities in the long range dilaton field of
any matter excitation of low energy two dimensional string theory
prevent us from concluding that the theory has the number of states
indicated by perturbation theory.  The naive picture of a moduli space
of vacua is undoubtedly also substantially altered, this time by quantum
fluctuations.  We have presented an argument that a proper treatment of
these fluctuations leads to the conclusion that string theory vacua with
one large spatial dimension were about as improbable is those with more
than four.  However, in high dimension we were justified in ignoring the
infrared quantum fluctuations of the gravitational field.  This is
likely to be incorrect in two dimensions, so our conclusions about the
vacuum state must be considered tentative.

\subsec{A Short Digression on Three Spacetime Dimensions}

It is tempting to try to extend these arguments to three spacetime
dimensions, since there too long range graviton and dilaton fields
prevent a straightforward perturbative treatment of the theory.
We believe however that the situation in three dimensions is quite
different.  Note that in this case, there is no violation of the
Bekenstein bound.

Let us begin by studying the low energy effective heterotic theory and consider
classical solutions involving only the gravitational, dilaton, and moduli
fields.  In the Einstein frame, the lagrangian has the form
\eqn\threedL{\sqrt{-g} (R - (\nabla\phi )^2 - G_{ij}(M)(\nabla_i
M)(\nabla_j M)).}
Notice that all couplings to the dilaton and other moduli are
derivative.  Thus, a smooth, small amplitude incoming wave
of massless matter will not produce logarithmically growing dilaton or
moduli fields in linear approximation.  This is no longer true if we
consider solutions involving vector fields, or if we consider
corrections to the field equations from higher orders in the string
tension expansion.   These will lead to nonderivative sources for the
dilaton.  In linear approximation,
the dilaton and other massless fields will grow
logarithmically at infinity.  One might expect such logarithmically
growing scalar fields to lead to infinite energies.

On the other hand, similar logarithmic fields are encountered in the
linearized approximation to the static fields of a massive particle.
We know however\ref\cosmic{B.R.Greene, A.Shapere, C.Vafa, S.T.Yau, {\it
Nucl. Phys.}{\bf B337},(1990),1; J.A.Bagger, C.G.Callan, J.A.Harvey,
{\it Nucl. Phys.}{\bf B278},(1986),550.} that four dimensional string theory
contains stringy cosmic strings with finite mass per unit length.  These
infinite straight
strings have a deficit angle at infinity which is substantial, but
less than $2\pi$ and
seem to be perfectly good, particlelike excitations of three dimensional
string theory.

It seems reasonable to conjecture then, that three dimensional string
vacua support a variety of states corresponding to finite numbers of
moving massive and massless particles.  If the particle number gets too
large, one of two things happens.  For systems (BPS saturated
multiparticle states would be one example) in which no bound states are
formed, the deficit angle will eventually exceed $2\pi$, and there will
be no large spacetime.  Alternatively, bound states can be formed of
arbitrary numbers of particles with deficit angle less than $2\pi$.
Thus, while string dynamics in three spacetime dimensions is
highly constrained, and quite odd, there does not seem to be any
lack of physical states.

The foregoing discussion applies to weakly coupled string theory.
For strong coupling, Witten\ref\ed{E.Witten, {\it Mod.Phys.Lett.}{\bf
A10},(1995),2153.} has suggested that at least some
three dimensional string vacua might turn out to look four dimensional.
In this case the infrared gravitational effects which make three
dimensional physics so bizarre would disappear at strong coupling, and
the theory would have many states.

\newsec{Conclusions}

The arguments of this paper suggest that classical string vacua with
geometries large compared to the string scale have a low probability of
being found in the quantum ground state, unless the large spacetime has
two, three or four dimensions.
Although the two dimensional case might lead to a satisfactory ground state for
string theory, it does not have any nearby excitations.
This is a satisfying conclusion because the naive picture of two
dimensional string theory appears to violate Bekenstein's entropy bound.
There is no longer an obvious counterargument to the claim
that string theory is a holographic theory\ref\holo{G. 't Hooft, {\it
Dimensional Reduction in Quantum Gravity}, in Salamfest 1993,
gr-qc/9310026;
L.Susskind,{J.Math.Phys.}{\bf 36}, (1995),6377.}.

Note that we do not claim to have solved the dynamics of two dimensional
string theory.  We merely showed that the heuristic perturbative picture
of the spectrum is incorrect.  We have also made a preliminary,
impressionistic study of the quantum ground state.  We argued that the
standard picture of a moduli space of vacua was incorrect.  On large
scales, the two dimensional observer will see a wave function that is
spread over moduli space.  Considerations of the dynamics of the
nonlinear sigma model on moduli space (but neglecting two dimensional
gravity, which may be a serious omission) suggest that the large scale
wave function
tends to be concentrated in noncompact regions of the moduli space.
However, precisely in those regions, the effective two dimensional field
theory breaks down because the space ceases to be two dimensional.
A better approximation to the dynamics is, we believe, obtained by
restoring the finite size of the last spatial dimension.  Then one might
argue that the two dimensional world is similar to that in dimensions
higher than $4$.  The vacuum wave function is concentrated in a region
where all spatial dimensions are of order the string scale.  Infrared
freedom of the sigma model on moduli space implies that no low energy
potential for the moduli is generated that could change this conclusion.
Furthermore, gauge dynamics in two dimensional string theory leads to
confinement on a scale of order the string scale, and should not alter
the conclusion either.  We are however much less certain of these
arguments than we are about the corresponding ones for high dimensions.
Two dimensional quantum gravity is not negligible in the infrared, and
we have neglected it in our considerations of the ground state.  Our
classical analysis of states containing arbitrarily small amounts of
matter suggests that this may be a serious error.

Although the dynamics of three spacetime dimensions
is very odd in string theory, as it will be in any theory containing
gravity and massless scalar fields, we could find no reason to expect a
restriction on the number of states in the theory.  Note that the three
dimensional theory does not lead to a violation of Bekenstein's principle.
The ultimate fate of three dimensional vacua of string theory is beyond
even our powers of speculation at this point.  It may be, as suggested
by Witten\ed , that some of these vacua are indeed $4$ dimensional
theories in disguise.
\listrefs
\end